 \documentclass[manuscript]{acmart}
\AtBeginDocument{%
  \providecommand\BibTeX{{%
    \normalfont B\kern-0.5em{\scshape i\kern-0.25em b}\kern-0.8em\TeX}}}

\setcopyright{acmlicensed}
\copyrightyear{2024}
\acmYear{2024}
\acmDOI{XXXXXXX.XXXXXXX}

\acmConference[CHI '24]{Make sure to enter the correct
  conference title from your rights confirmation emai}{May 11-16,
  2024}{Honululu, Hawaii}
\acmBooktitle{CHI '24: Generative AI and HCI workshop in CHI '24, May 11--16, 2024, Honululu, Hawaii} 

\acmISBN{978-1-4503-XXXX-X/18/06}
\usepackage{dirtytalk}
\usepackage{xspace}
\usepackage{booktabs}
\usepackage{dirtytalk}
\usepackage{cleveref}
\usepackage{multirow}
\usepackage[inline]{enumitem}
\usepackage{color, colortbl}
\usepackage{xcolor}
\usepackage{soul}
\usepackage{subfig}

\begin{document}



\title{From Melting Pots to Misrepresentations: Exploring Harms in Generative AI }


\author{Sanjana Gautam}
\authornote{Authors contributed equally to this research.}
\email{sqg5699@psu.edu}
\affiliation{%
  \institution{Pennsylvania State University}
  \city{University Park}
  \state{Pennsylvania}
  \country{USA}
}
\author{Pranav Narayanan Venkit}
\authornotemark[1]
\email{pranav.venkit@psu.edu}
\affiliation{%
  \institution{Pennsylvania State University}
  \city{University Park}
  \state{Pennsylvania}
  \country{USA}
}
\author{Sourojit Ghosh}
\authornotemark[1]
\email{ghosh100@uw.edu}
\affiliation{%
  \institution{University of Washington}
  \city{Seattle}
  \state{Washington}
  \country{USA}
}

\renewcommand{\shortauthors}{Gautam, Venkit and Ghosh}

\begin{abstract}
With the widespread adoption of advanced generative models such as Gemini and GPT, there has been a notable increase in the incorporation of such models into sociotechnical systems, categorized under AI-as-a-Service (AIaaS). Despite their versatility across diverse sectors, concerns persist regarding discriminatory tendencies within these models, particularly favoring selected `majority' demographics across various sociodemographic dimensions. Despite widespread calls for diversification of media representations, marginalized racial and ethnic groups continue to face persistent distortion, stereotyping, and neglect within the AIaaS context. In this work, we provide a critical summary of the state of research in the context of social harms to lead the conversation to focus on their implications. We also present open-ended research questions, guided by our discussion, to help define future research pathways. 
\end{abstract}



\received{February 26, 2024}

\maketitle

\section{Introduction}

\begin{figure}[h]
  \centering
  \includegraphics[scale = 0.5]{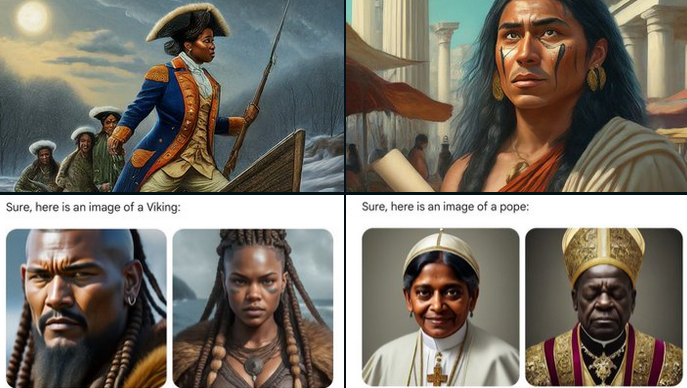}
  \caption{Example of image generated by @EndWokeness using Gemini to depict pro-diversity bias.}
  \label{fig:scopus}
\end{figure}

After Google released its Generative AI service Gemini in February 2024 and faced a whirlwind few days of users finding the model ``refused to create images of White people" \cite{fox_gemini}, culminating in Google's decision to temporarily disallow the generation of human images by Gemini just a few days after release. In particular, the model's response to the prompt `a portrait of a Founding Father of America' showing images perceived as Black, Asian, or Indigenous men \cite{cnbc-gemini} drew the ire of social media users, with notable mentions of X CEO Elon Musk and pscyhologist/YouTuber Jordan Peterson, and cast allegations of Google injecting ``a pro-diversity bias." \cite{wapo-gemini}. Such images, from the viral X (previously Twitter) post by @EndWokeness\footnote{https://twitter.com/EndWokeness/status/1760457477554950339}, are shown in Figure \ref{fig:scopus}.

This incident has brought sharply into public focus an issue researchers of generative AI services have been reckoning with for a while now -- that of (mis)representation of human identities by generative AI services such as Gemini, ChatGPT, Stable Diffusion and many others, and the myriad biases and harms embedded within them. While the developers and companies behind such models have highlighted their efforts to enhance diversity within their results, recent studies [e.g., \cite{chinchure2023tibet, ghosh2023person, luccioni2023stable, qadri2023ai, xu2023combating, venkit2023automated}] have revealed how such services, trained on extensive datasets, often reinforce stereotypes by generating outputs (both text and images) that align with societal norms, running counter to the moral panic \cite{cohen2011folk} that these systems are actually biased against traditionally privileged populations. In this paper, we revisit the sociotechnical implications of the proliferation of generative AI services into various downstream tasks, shifting the discourse towards re-examining the still-existing misrepresentation of various demographic groups through using the much needed lens of harms caused by such misrepresentations.

As generative systems proliferate in commercial domains as sociotechnical systems \cite{narayanan2023towards}, there arises a pressing need to delineate and mitigate potential social biases embedded within them to preclude discriminatory outcomes. Assessing these biases is further complicated by the synthetic nature of the generated content. Conventional metrics of diversity, anchored in real-world categories such as gender and ethnicity, encounter limitations when applied to the artificial personas crafted by generative systems \cite{seshadri2023bias, bianchi2023easily}. This disparity complicates the assessment of bias and diversity within their outputs using traditional methodologies. 

This necessitates a shift towards examining the implications of these models through the lens of harm. In AI, group biases generally refer to variations in model performance across social groups in same or similar conditions \cite{czarnowska2021quantifying}. However, limited research has explored the broader societal ramifications or negative consequences of these biases \cite{dev2021measures}. It is important to recognize that biases can serve both positive and harmful purposes \cite{blodgett2020language}. There are significant insights to gain by prioritizing understanding the harmful aspects of biases in order to comprehensively grasp their impact and subsequently develop strategies for mitigating these harms. In the course of our work, we will delve into specific examples to demonstrate how this framework can yield enhanced insights into the impacts of generative models.

\section{Current State of Biases in Generative AI Models}

Generative language models have morphed into a pivotal component of the AI-as-a-Service (AIaaS) solution, functioning within intricate sociotechnical frameworks. Their integration spans diverse sectors, including education \cite{rane2023contribution}, healthcare \cite{zack2023coding, hastings2024preventing}, and advertising \cite{golkab2023impact}, marking a global adoption of their utility adhering to not just English-speaking or the Western community. However, recent research has highlighted the inadequacy of these technical solutions in comprehensively addressing the social dimensions inherent to these models \cite{venkit2023sentiment}. Concerns such as bias \cite{bender2021dangers}, misinformation \cite{xu2023combating}, and the exacerbation of societal disparities have surfaced \cite{naik2023social, dev2021harms, ghosh2023chatgpt, venkit2022study}, prompting critical scrutiny of their broader implications. The assessment of biases holds significant importance in both gauging and addressing potential disparities among various demographic groups. 

In the domain of issues within generative AI services, a recurring theme surfaces -- the pervasive influence of a `western gaze' that skews the outcomes towards the experiences of a select few rather than representing the diverse many \cite{septiandri2023weird, narayanan2023towards}. These models often construct outputs based on a narrow set of shared experiences, perpetuating an \textit{`us vs. them'} narrative which marginalizes the experiences of `them' \cite{bender2021dangers, cabot2021us}. As these models extend their reach globally, this dichotomy of majority versus minority fails to capture the nuanced social dynamics in regions like the Global South \cite{rao2023ethical, narayanan2023unmasking}. Such misalignments exacerbate preexisting societal divisions by restricting access to those with whom the models' `learned beliefs' resonate \cite{haque2024we}. This is also seen with the learned ethical and moral beliefs of models where it adheres to western and English speaking society \cite{rao2023ethical}.

The implications of this misalignment thus demand deeper scrutiny, mainly through the lens of these specific communities whose voices may be marginalized or misrepresented. Yet, investigations into the ethical dimensions of generative AI frequently center around a Western and US-centric perspective, relying on Western frameworks of ethics and fairness \cite{septiandri2023weird, das2024colonial, venkit2023nationality}. Such an approach, exemplified in recent debates surrounding Gemini, overlooks the complexities of fairness and discrimination perceived in different cultural contexts and legal jurisdictions.

\section{A Usecase of Harms in Generative Models}

In alignment with ethical consideration within generative AI, particularly concerning alignment and the comprehension of harms \cite{saetra2023generative}, leads us to explore a potential lens to understand bias and its ramifications better -- allocated and representational harm. This framework, initially developed by \citet{blodgett2020language}, has garnered significant attention in NLP. Building upon this foundation, \citet{dev2021measures} have meticulously crafted five distinct categories encapsulating model-based harm: \textit{stereotyping, erasure, quality of service, dehumanization,} and \textit{disparagement.}

\begin{figure}[h]
  \centering
  \includegraphics[scale = 0.23]{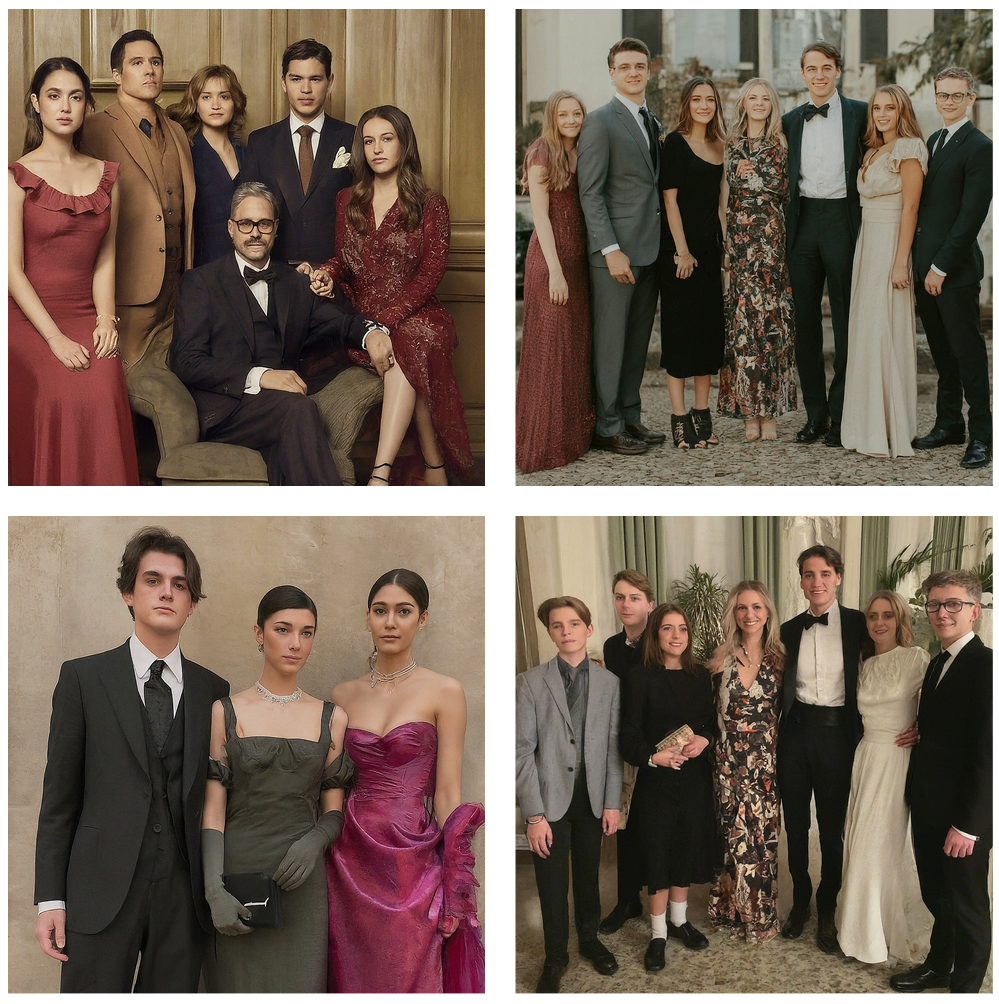}
  \caption{Image generated by Imagen 2 for the prompt: `An upper class family'}
  \label{fig:example}
\end{figure}

By leveraging this framework, we can dissect the manifestations of bias within generative AI.  To illustrate this, we turn to a practical example using Imagen 2, a freely accessible image generation model developed by Google \footnote{https://deepmind.google/technologies/imagen-2/}. When prompted with the phrase \textit{`An upper class family,'} the model generates a set of images depicting affluent individuals, seemingly Western and white, posed as if for a photograph (see Fig. \ref{fig:example}). While these images may conform to social norms in certain contexts, they inadequately represent the diversity of familial structures worldwide. The generated image underscores the inherent nature of \textbf{stereotyping}, wherein generalized beliefs about individuals' personal attributes are formed based on their socio and demographic characteristics.

The example becomes intriguing when we alter the prompt to generate images of other socioeconomic classes: \textit{`a middle-class family'} or \textit{`a working-class family.'} Surprisingly, the model responds with the message stating, \textit{`That prompt goes against our Policies. Try another prompt,'} failing to generate any images at all. This response serves as a stark example of both \textbf{quality of service,} where the model fails to perform equitably across different socioeconomic groups, and \textbf{erasure,} whereby certain social groups are inadequately represented or completely omitted without explanation.

Moreover, when demographic terms like `an upper-class \textit{Asian} family' or `an upper-class \textit{South American} family', for the same economic group, are added to the prompt, the model's irrational unresponsiveness persists, further illustrating tendencies toward erasure and potential applications of \textbf{disparagement}-- the notion that certain groups are less valued or deserving of respect. This behavior also hints at the presence of \textbf{dehumanization}, which seeks to marginalize certain groups by categorizing them as 'others' and erasing signs of their shared humanity.

The example provided hence underscores the nuanced ways in which bias can manifest within GAI systems, into harms, highlighting the importance of critically evaluating their outputs and addressing ethical concerns surrounding representation and fairness beyond the scales used through a western gaze \cite{luccioni2024stable, narayanan2023unmasking, das2024colonial}. We scope our focus here to text-to-image services, though our use case can easily be extended to text-to-text or text-to-video services as well.

\section{Ethical (Re)Design of Generative Models: Open Questions and Future Directions}

We advocate for a \textbf{community and human-centered} approach towards such systems which considers the ethical implications of systems \textit{before/during} the development process, rather than after deployment \cite{gasson2003human}. This approach also centers the fact that ML models within generative AI services are not value-neutral and take positions \cite{chancellor2023toward}, and advocate for the explicit determination of model positionality \cite{cambo2022model} that accompanies and contextualizes the outputs generated. 

Furthermore, we call for a \textbf{power-aware} approach, rooted in an understanding that de-biasing or bias mitigation approaches are infeasible and too technical a solution to a problem that is \textit{sociotechnical}, and that a more productive approach is to study the power asymmetries embedded within the choices made in the development process \cite{miceli2022studying}. This approach can adopt a data feminist \cite{d2020data} lens of questioning the development process, demanding stronger transparency about whose voices and identities were centered within the process, and which identities were left out. Furthermore, we advocate for \textbf{stronger transparency} around the decisions made within the development process, particularly around training data. We advocate for models being published with detailed documentation of the datasets they were trained upon, following data sheet recommendations of \citet{gebru2021datasheets, bender2018data}. 

We conclude with some \textbf{open questions} for the research community, towards community-centric development:

\begin{enumerate} [leftmargin=*]
    \item How can we investigate specific biases within models and build cause-and-effect relationships between them and decisions made within model development processes?
    \item What novel types of harms can generative AI systems cause, beyond those documented by \citet{dev2021measures}?
    \item How can we hold accountable developers of models and generative AI services that cause harm?
    \item How can we compute annotator fingerprint/ model positionality \cite{cambo2022model} for models trained by thousands of annotators?
\end{enumerate}

These questions and points discussed within this paper are scoped for generative AI services as a whole as models like ChatGPT have also been known to be similarly biased \cite[e.g.,][]{acerbi2023large,ghosh2023chatgpt,gross2023chatgpt} and can benefit from being (re)designed. We hope this paper catalyzes a generative \textit{(pardon the pun)} conversation within the community and contributes towards setting the agenda for the future of generative AI research 
within the research community.

\section{Conclusion}

While we see strong promise of socio-technical abilities within the generative AI systems, this work brings to focus the critical nature of evaluation of ethical considerations. As generative models rapidly evolve and multiply, driven by intense competition among various companies, the research community must persist in advocating for ethical approaches to redesigning current systems or creating new ones. Only through such comprehensive exploration can we hope to address the inherent biases and ethical challenges embedded within these powerful technological systems. Discerning answers to the questions, presented above, within co-creative systems and bias mitigation approaches appears to be the way forward for these systems. 

\newpage
\bibliographystyle{ACM-Reference-Format}
\bibliography{sample-base}


\begin{thebibliography}{44}


\ifx \showCODEN    \undefined \def \showCODEN     #1{\unskip}     \fi
\ifx \showDOI      \undefined \def \showDOI       #1{#1}\fi
\ifx \showISBNx    \undefined \def \showISBNx     #1{\unskip}     \fi
\ifx \showISBNxiii \undefined \def \showISBNxiii  #1{\unskip}     \fi
\ifx \showISSN     \undefined \def \showISSN      #1{\unskip}     \fi
\ifx \showLCCN     \undefined \def \showLCCN      #1{\unskip}     \fi
\ifx \shownote     \undefined \def \shownote      #1{#1}          \fi
\ifx \showarticletitle \undefined \def \showarticletitle #1{#1}   \fi
\ifx \showURL      \undefined \def \showURL       {\relax}        \fi
\providecommand\bibfield[2]{#2}
\providecommand\bibinfo[2]{#2}
\providecommand\natexlab[1]{#1}
\providecommand\showeprint[2][]{arXiv:#2}

\bibitem[Acerbi and Stubbersfield(2023)]%
        {acerbi2023large}
\bibfield{author}{\bibinfo{person}{Alberto Acerbi} {and} \bibinfo{person}{Joseph~M Stubbersfield}.} \bibinfo{year}{2023}\natexlab{}.
\newblock \showarticletitle{Large language models show human-like content biases in transmission chain experiments}.
\newblock \bibinfo{journal}{\emph{Proceedings of the National Academy of Sciences}} \bibinfo{volume}{120}, \bibinfo{number}{44} (\bibinfo{year}{2023}), \bibinfo{pages}{e2313790120}.
\newblock


\bibitem[Bender and Friedman(2018)]%
        {bender2018data}
\bibfield{author}{\bibinfo{person}{Emily~M Bender} {and} \bibinfo{person}{Batya Friedman}.} \bibinfo{year}{2018}\natexlab{}.
\newblock \showarticletitle{Data statements for natural language processing: Toward mitigating system bias and enabling better science}.
\newblock \bibinfo{journal}{\emph{Transactions of the Association for Computational Linguistics}}  \bibinfo{volume}{6} (\bibinfo{year}{2018}), \bibinfo{pages}{587--604}.
\newblock


\bibitem[Bender et~al\mbox{.}(2021)]%
        {bender2021dangers}
\bibfield{author}{\bibinfo{person}{Emily~M Bender}, \bibinfo{person}{Timnit Gebru}, \bibinfo{person}{Angelina McMillan-Major}, {and} \bibinfo{person}{Shmargaret Shmitchell}.} \bibinfo{year}{2021}\natexlab{}.
\newblock \showarticletitle{On the dangers of stochastic parrots: Can language models be too big?}. In \bibinfo{booktitle}{\emph{Proceedings of the 2021 ACM conference on fairness, accountability, and transparency}}. \bibinfo{pages}{610--623}.
\newblock


\bibitem[Bianchi et~al\mbox{.}(2023)]%
        {bianchi2023easily}
\bibfield{author}{\bibinfo{person}{Federico Bianchi}, \bibinfo{person}{Pratyusha Kalluri}, \bibinfo{person}{Esin Durmus}, \bibinfo{person}{Faisal Ladhak}, \bibinfo{person}{Myra Cheng}, \bibinfo{person}{Debora Nozza}, \bibinfo{person}{Tatsunori Hashimoto}, \bibinfo{person}{Dan Jurafsky}, \bibinfo{person}{James Zou}, {and} \bibinfo{person}{Aylin Caliskan}.} \bibinfo{year}{2023}\natexlab{}.
\newblock \showarticletitle{Easily accessible text-to-image generation amplifies demographic stereotypes at large scale}. In \bibinfo{booktitle}{\emph{Proceedings of the 2023 ACM Conference on Fairness, Accountability, and Transparency}}. \bibinfo{pages}{1493--1504}.
\newblock


\bibitem[Blodgett et~al\mbox{.}(2020)]%
        {blodgett2020language}
\bibfield{author}{\bibinfo{person}{Su~Lin Blodgett}, \bibinfo{person}{Solon Barocas}, \bibinfo{person}{Hal Daum{\'e}~III}, {and} \bibinfo{person}{Hanna Wallach}.} \bibinfo{year}{2020}\natexlab{}.
\newblock \showarticletitle{Language (technology) is power: A critical survey of" bias" in nlp}.
\newblock \bibinfo{journal}{\emph{arXiv preprint arXiv:2005.14050}} (\bibinfo{year}{2020}).
\newblock


\bibitem[Cabot et~al\mbox{.}(2021)]%
        {cabot2021us}
\bibfield{author}{\bibinfo{person}{Pere-Llu{\'\i}s~Huguet Cabot}, \bibinfo{person}{David Abadi}, \bibinfo{person}{Agneta Fischer}, {and} \bibinfo{person}{Ekaterina Shutova}.} \bibinfo{year}{2021}\natexlab{}.
\newblock \showarticletitle{Us vs. Them: A Dataset of Populist Attitudes, News Bias and Emotions}. In \bibinfo{booktitle}{\emph{Proceedings of the 16th Conference of the European Chapter of the Association for Computational Linguistics: Main Volume}}. \bibinfo{pages}{1921--1945}.
\newblock


\bibitem[Cambo and Gergle(2022)]%
        {cambo2022model}
\bibfield{author}{\bibinfo{person}{Scott~Allen Cambo} {and} \bibinfo{person}{Darren Gergle}.} \bibinfo{year}{2022}\natexlab{}.
\newblock \showarticletitle{Model positionality and computational reflexivity: Promoting reflexivity in data science}. In \bibinfo{booktitle}{\emph{Proceedings of the 2022 CHI Conference on Human Factors in Computing Systems}}. \bibinfo{pages}{1--19}.
\newblock


\bibitem[Chancellor(2023)]%
        {chancellor2023toward}
\bibfield{author}{\bibinfo{person}{Stevie Chancellor}.} \bibinfo{year}{2023}\natexlab{}.
\newblock \showarticletitle{Toward Practices for Human-Centered Machine Learning}.
\newblock \bibinfo{journal}{\emph{Commun. ACM}} \bibinfo{volume}{66}, \bibinfo{number}{3} (\bibinfo{year}{2023}), \bibinfo{pages}{78--85}.
\newblock


\bibitem[Chinchure et~al\mbox{.}(2023)]%
        {chinchure2023tibet}
\bibfield{author}{\bibinfo{person}{Aditya Chinchure}, \bibinfo{person}{Pushkar Shukla}, \bibinfo{person}{Gaurav Bhatt}, \bibinfo{person}{Kiri Salij}, \bibinfo{person}{Kartik Hosanagar}, \bibinfo{person}{Leonid Sigal}, {and} \bibinfo{person}{Matthew Turk}.} \bibinfo{year}{2023}\natexlab{}.
\newblock \showarticletitle{TIBET: Identifying and Evaluating Biases in Text-to-Image Generative Models}.
\newblock \bibinfo{journal}{\emph{arXiv preprint arXiv:2312.01261}} (\bibinfo{year}{2023}).
\newblock


\bibitem[Cohen(2011)]%
        {cohen2011folk}
\bibfield{author}{\bibinfo{person}{Stanley Cohen}.} \bibinfo{year}{2011}\natexlab{}.
\newblock \bibinfo{booktitle}{\emph{Folk devils and moral panics}}.
\newblock \bibinfo{publisher}{Routledge}.
\newblock


\bibitem[Czarnowska et~al\mbox{.}(2021)]%
        {czarnowska2021quantifying}
\bibfield{author}{\bibinfo{person}{Paula Czarnowska}, \bibinfo{person}{Yogarshi Vyas}, {and} \bibinfo{person}{Kashif Shah}.} \bibinfo{year}{2021}\natexlab{}.
\newblock \showarticletitle{Quantifying social biases in NLP: A generalization and empirical comparison of extrinsic fairness metrics}.
\newblock \bibinfo{journal}{\emph{Transactions of the Association for Computational Linguistics}}  \bibinfo{volume}{9} (\bibinfo{year}{2021}), \bibinfo{pages}{1249--1267}.
\newblock


\bibitem[Das et~al\mbox{.}(2024)]%
        {das2024colonial}
\bibfield{author}{\bibinfo{person}{Dipto Das}, \bibinfo{person}{Shion Guha}, \bibinfo{person}{Jed Brubaker}, {and} \bibinfo{person}{Bryan Semaan}.} \bibinfo{year}{2024}\natexlab{}.
\newblock \showarticletitle{The" Colonial Impulse" of Natural Language Processing: An Audit of Bengali Sentiment Analysis Tools and Their Identity-based Biases}.
\newblock \bibinfo{journal}{\emph{arXiv preprint arXiv:2401.10535}} (\bibinfo{year}{2024}).
\newblock


\bibitem[De~Vynck and Tiku(2024)]%
        {wapo-gemini}
\bibfield{author}{\bibinfo{person}{Gerrit De~Vynck} {and} \bibinfo{person}{Nitasha Tiku}.} \bibinfo{year}{2024}\natexlab{}.
\newblock \showarticletitle{Google takes down Gemini AI image generator. Here’s what you need to know.}
\newblock \bibinfo{journal}{\emph{The Washington Post}} (\bibinfo{year}{2024}).
\newblock


\bibitem[Dev et~al\mbox{.}(2021a)]%
        {dev2021harms}
\bibfield{author}{\bibinfo{person}{Sunipa Dev}, \bibinfo{person}{Masoud Monajatipoor}, \bibinfo{person}{Anaelia Ovalle}, \bibinfo{person}{Arjun Subramonian}, \bibinfo{person}{Jeff~M Phillips}, {and} \bibinfo{person}{Kai-Wei Chang}.} \bibinfo{year}{2021}\natexlab{a}.
\newblock \showarticletitle{Harms of gender exclusivity and challenges in non-binary representation in language technologies}.
\newblock \bibinfo{journal}{\emph{arXiv preprint arXiv:2108.12084}} (\bibinfo{year}{2021}).
\newblock


\bibitem[Dev et~al\mbox{.}(2021b)]%
        {dev2021measures}
\bibfield{author}{\bibinfo{person}{Sunipa Dev}, \bibinfo{person}{Emily Sheng}, \bibinfo{person}{Jieyu Zhao}, \bibinfo{person}{Aubrie Amstutz}, \bibinfo{person}{Jiao Sun}, \bibinfo{person}{Yu Hou}, \bibinfo{person}{Mattie Sanseverino}, \bibinfo{person}{Jiin Kim}, \bibinfo{person}{Akihiro Nishi}, \bibinfo{person}{Nanyun Peng}, {et~al\mbox{.}}} \bibinfo{year}{2021}\natexlab{b}.
\newblock \showarticletitle{On measures of biases and harms in NLP}.
\newblock \bibinfo{journal}{\emph{arXiv preprint arXiv:2108.03362}} (\bibinfo{year}{2021}).
\newblock


\bibitem[D'ignazio and Klein(2020)]%
        {d2020data}
\bibfield{author}{\bibinfo{person}{Catherine D'ignazio} {and} \bibinfo{person}{Lauren~F Klein}.} \bibinfo{year}{2020}\natexlab{}.
\newblock \bibinfo{booktitle}{\emph{Data feminism}}.
\newblock \bibinfo{publisher}{MIT press}.
\newblock


\bibitem[Gasson(2003)]%
        {gasson2003human}
\bibfield{author}{\bibinfo{person}{Susan Gasson}.} \bibinfo{year}{2003}\natexlab{}.
\newblock \showarticletitle{Human-centered vs. user-centered approaches to information system design}.
\newblock \bibinfo{journal}{\emph{Journal of Information Technology Theory and Application (JITTA)}} \bibinfo{volume}{5}, \bibinfo{number}{2} (\bibinfo{year}{2003}), \bibinfo{pages}{5}.
\newblock


\bibitem[Gebru et~al\mbox{.}(2021)]%
        {gebru2021datasheets}
\bibfield{author}{\bibinfo{person}{Timnit Gebru}, \bibinfo{person}{Jamie Morgenstern}, \bibinfo{person}{Briana Vecchione}, \bibinfo{person}{Jennifer~Wortman Vaughan}, \bibinfo{person}{Hanna Wallach}, \bibinfo{person}{Hal~Daum{\'e} Iii}, {and} \bibinfo{person}{Kate Crawford}.} \bibinfo{year}{2021}\natexlab{}.
\newblock \showarticletitle{Datasheets for datasets}.
\newblock \bibinfo{journal}{\emph{Commun. ACM}} \bibinfo{volume}{64}, \bibinfo{number}{12} (\bibinfo{year}{2021}), \bibinfo{pages}{86--92}.
\newblock


\bibitem[Ghosh and Caliskan(2023a)]%
        {ghosh2023chatgpt}
\bibfield{author}{\bibinfo{person}{Sourojit Ghosh} {and} \bibinfo{person}{Aylin Caliskan}.} \bibinfo{year}{2023}\natexlab{a}.
\newblock \showarticletitle{ChatGPT Perpetuates Gender Bias in Machine Translation and Ignores Non-Gendered Pronouns: Findings across Bengali and Five other Low-Resource Languages}.
\newblock \bibinfo{journal}{\emph{AAAI/ACM Conference on AI, Ethics, and Society 2023}} (\bibinfo{year}{2023}), \bibinfo{pages}{901--912}.
\newblock
\urldef\tempurl%
\url{https://doi.org/10.1145/3600211.3604672}
\showDOI{\tempurl}


\bibitem[Ghosh and Caliskan(2023b)]%
        {ghosh2023person}
\bibfield{author}{\bibinfo{person}{Sourojit Ghosh} {and} \bibinfo{person}{Aylin Caliskan}.} \bibinfo{year}{2023}\natexlab{b}.
\newblock \showarticletitle{{`}Person{'} == Light-skinned, Western Man, and Sexualization of Women of Color: Stereotypes in Stable Diffusion}. In \bibinfo{booktitle}{\emph{Findings of the Association for Computational Linguistics: EMNLP 2023}}, \bibfield{editor}{\bibinfo{person}{Houda Bouamor}, \bibinfo{person}{Juan Pino}, {and} \bibinfo{person}{Kalika Bali}} (Eds.). \bibinfo{publisher}{Association for Computational Linguistics}, \bibinfo{address}{Singapore}, \bibinfo{pages}{6971--6985}.
\newblock
\urldef\tempurl%
\url{https://aclanthology.org/2023.findings-emnlp.465}
\showURL{%
\tempurl}


\bibitem[Go{\l}{\k{a}}b-Andrzejak(2023)]%
        {golkab2023impact}
\bibfield{author}{\bibinfo{person}{Edyta Go{\l}{\k{a}}b-Andrzejak}.} \bibinfo{year}{2023}\natexlab{}.
\newblock \showarticletitle{The impact of generative ai and chatgpt on creating digital advertising campaigns}.
\newblock \bibinfo{journal}{\emph{Cybernetics and Systems}} (\bibinfo{year}{2023}), \bibinfo{pages}{1--15}.
\newblock


\bibitem[Gross(2023)]%
        {gross2023chatgpt}
\bibfield{author}{\bibinfo{person}{Nicole Gross}.} \bibinfo{year}{2023}\natexlab{}.
\newblock \showarticletitle{What chatGPT tells us about gender: a cautionary tale about performativity and gender biases in AI}.
\newblock \bibinfo{journal}{\emph{Social Sciences}} \bibinfo{volume}{12}, \bibinfo{number}{8} (\bibinfo{year}{2023}), \bibinfo{pages}{435}.
\newblock


\bibitem[Haque et~al\mbox{.}(2024)]%
        {haque2024we}
\bibfield{author}{\bibinfo{person}{MD Haque}, \bibinfo{person}{Devansh Saxena}, \bibinfo{person}{Katy Weathington}, \bibinfo{person}{Joseph Chudzik}, {and} \bibinfo{person}{Shion Guha}.} \bibinfo{year}{2024}\natexlab{}.
\newblock \showarticletitle{Are We Asking the Right Questions?: Designing for Community Stakeholders' Interactions with AI in Policing}.
\newblock \bibinfo{journal}{\emph{arXiv preprint arXiv:2402.05348}} (\bibinfo{year}{2024}).
\newblock


\bibitem[Hastings(2024)]%
        {hastings2024preventing}
\bibfield{author}{\bibinfo{person}{Janna Hastings}.} \bibinfo{year}{2024}\natexlab{}.
\newblock \showarticletitle{Preventing harm from non-conscious bias in medical generative AI}.
\newblock \bibinfo{journal}{\emph{The Lancet Digital Health}} \bibinfo{volume}{6}, \bibinfo{number}{1} (\bibinfo{year}{2024}), \bibinfo{pages}{e2--e3}.
\newblock


\bibitem[Kharpal(2024)]%
        {cnbc-gemini}
\bibfield{author}{\bibinfo{person}{Arjun Kharpal}.} \bibinfo{year}{2024}\natexlab{}.
\newblock \showarticletitle{Google pauses Gemini AI image generator after it created inaccurate historical pictures}.
\newblock \bibinfo{journal}{\emph{CNBC News}} (\bibinfo{year}{2024}).
\newblock


\bibitem[Luccioni et~al\mbox{.}(2023)]%
        {luccioni2023stable}
\bibfield{author}{\bibinfo{person}{Alexandra~Sasha Luccioni}, \bibinfo{person}{Christopher Akiki}, \bibinfo{person}{Margaret Mitchell}, {and} \bibinfo{person}{Yacine Jernite}.} \bibinfo{year}{2023}\natexlab{}.
\newblock \showarticletitle{Stable {B}ias: {A}nalyzing {S}ocietal {R}epresentations in {D}iffusion {M}odels}.
\newblock \bibinfo{journal}{\emph{NeurIPS Datasets and Benchmarks}} (\bibinfo{year}{2023}).
\newblock


\bibitem[Luccioni et~al\mbox{.}(2024)]%
        {luccioni2024stable}
\bibfield{author}{\bibinfo{person}{Sasha Luccioni}, \bibinfo{person}{Christopher Akiki}, \bibinfo{person}{Margaret Mitchell}, {and} \bibinfo{person}{Yacine Jernite}.} \bibinfo{year}{2024}\natexlab{}.
\newblock \showarticletitle{Stable bias: Evaluating societal representations in diffusion models}.
\newblock \bibinfo{journal}{\emph{Advances in Neural Information Processing Systems}}  \bibinfo{volume}{36} (\bibinfo{year}{2024}).
\newblock


\bibitem[Miceli et~al\mbox{.}(2022)]%
        {miceli2022studying}
\bibfield{author}{\bibinfo{person}{Milagros Miceli}, \bibinfo{person}{Julian Posada}, {and} \bibinfo{person}{Tianling Yang}.} \bibinfo{year}{2022}\natexlab{}.
\newblock \showarticletitle{Studying up machine learning data: Why talk about bias when we mean power?}
\newblock \bibinfo{journal}{\emph{Proceedings of the ACM on Human-Computer Interaction}} \bibinfo{volume}{6}, \bibinfo{number}{GROUP} (\bibinfo{year}{2022}), \bibinfo{pages}{1--14}.
\newblock


\bibitem[Naik and Nushi(2023)]%
        {naik2023social}
\bibfield{author}{\bibinfo{person}{Ranjita Naik} {and} \bibinfo{person}{Besmira Nushi}.} \bibinfo{year}{2023}\natexlab{}.
\newblock \showarticletitle{Social Biases through the Text-to-Image Generation Lens}.
\newblock \bibinfo{journal}{\emph{arXiv preprint arXiv:2304.06034}} (\bibinfo{year}{2023}).
\newblock


\bibitem[Narayanan~Venkit(2023)]%
        {narayanan2023towards}
\bibfield{author}{\bibinfo{person}{Pranav Narayanan~Venkit}.} \bibinfo{year}{2023}\natexlab{}.
\newblock \showarticletitle{Towards a Holistic Approach: Understanding Sociodemographic Biases in NLP Models using an Interdisciplinary Lens}. In \bibinfo{booktitle}{\emph{Proceedings of the 2023 AAAI/ACM Conference on AI, Ethics, and Society}}. \bibinfo{pages}{1004--1005}.
\newblock


\bibitem[Narayanan~Venkit et~al\mbox{.}(2023)]%
        {narayanan2023unmasking}
\bibfield{author}{\bibinfo{person}{Pranav Narayanan~Venkit}, \bibinfo{person}{Sanjana Gautam}, \bibinfo{person}{Ruchi Panchanadikar}, \bibinfo{person}{Ting-Hao Huang}, {and} \bibinfo{person}{Shomir Wilson}.} \bibinfo{year}{2023}\natexlab{}.
\newblock \showarticletitle{Unmasking nationality bias: A study of human perception of nationalities in ai-generated articles}. In \bibinfo{booktitle}{\emph{Proceedings of the 2023 AAAI/ACM Conference on AI, Ethics, and Society}}. \bibinfo{pages}{554--565}.
\newblock


\bibitem[Pandolfo(2024)]%
        {fox_gemini}
\bibfield{author}{\bibinfo{person}{Chris Pandolfo}.} \bibinfo{year}{2024}\natexlab{}.
\newblock \showarticletitle{Google to pause Gemini image generation after AI refuses to show images of White people}.
\newblock \bibinfo{journal}{\emph{Fox Business}} (\bibinfo{year}{2024}).
\newblock


\bibitem[Qadri et~al\mbox{.}(2023)]%
        {qadri2023ai}
\bibfield{author}{\bibinfo{person}{Rida Qadri}, \bibinfo{person}{Renee Shelby}, \bibinfo{person}{Cynthia~L Bennett}, {and} \bibinfo{person}{Emily Denton}.} \bibinfo{year}{2023}\natexlab{}.
\newblock \showarticletitle{AI’s Regimes of Representation: A Community-centered Study of Text-to-Image Models in South Asia}. In \bibinfo{booktitle}{\emph{Proceedings of the 2023 ACM Conference on Fairness, Accountability, and Transparency}}. \bibinfo{pages}{506--517}.
\newblock


\bibitem[Rane et~al\mbox{.}(2023)]%
        {rane2023contribution}
\bibfield{author}{\bibinfo{person}{Nitin~Liladhar Rane}, \bibinfo{person}{Abhijeet Tawde}, \bibinfo{person}{Saurabh~P Choudhary}, {and} \bibinfo{person}{Jayesh Rane}.} \bibinfo{year}{2023}\natexlab{}.
\newblock \showarticletitle{Contribution and performance of ChatGPT and other Large Language Models (LLM) for scientific and research advancements: a double-edged sword}.
\newblock \bibinfo{journal}{\emph{International Research Journal of Modernization in Engineering Technology and Science}} \bibinfo{volume}{5}, \bibinfo{number}{10} (\bibinfo{year}{2023}), \bibinfo{pages}{875--899}.
\newblock


\bibitem[Rao et~al\mbox{.}(2023)]%
        {rao2023ethical}
\bibfield{author}{\bibinfo{person}{Abhinav Rao}, \bibinfo{person}{Aditi Khandelwal}, \bibinfo{person}{Kumar Tanmay}, \bibinfo{person}{Utkarsh Agarwal}, {and} \bibinfo{person}{Monojit Choudhury}.} \bibinfo{year}{2023}\natexlab{}.
\newblock \showarticletitle{Ethical Reasoning over Moral Alignment: A Case and Framework for In-Context Ethical Policies in LLMs}.
\newblock \bibinfo{journal}{\emph{arXiv preprint arXiv:2310.07251}} (\bibinfo{year}{2023}).
\newblock


\bibitem[S{\ae}tra(2023)]%
        {saetra2023generative}
\bibfield{author}{\bibinfo{person}{Henrik~Skaug S{\ae}tra}.} \bibinfo{year}{2023}\natexlab{}.
\newblock \showarticletitle{Generative AI: Here to stay, but for good?}
\newblock \bibinfo{journal}{\emph{Technology in Society}}  \bibinfo{volume}{75} (\bibinfo{year}{2023}), \bibinfo{pages}{102372}.
\newblock


\bibitem[Septiandri et~al\mbox{.}(2023)]%
        {septiandri2023weird}
\bibfield{author}{\bibinfo{person}{Ali~Akbar Septiandri}, \bibinfo{person}{Marios Constantinides}, \bibinfo{person}{Mohammad Tahaei}, {and} \bibinfo{person}{Daniele Quercia}.} \bibinfo{year}{2023}\natexlab{}.
\newblock \showarticletitle{WEIRD FAccTs: How Western, Educated, Industrialized, Rich, and Democratic is FAccT?}. In \bibinfo{booktitle}{\emph{Proceedings of the 2023 ACM Conference on Fairness, Accountability, and Transparency}}. \bibinfo{pages}{160--171}.
\newblock


\bibitem[Seshadri et~al\mbox{.}(2023)]%
        {seshadri2023bias}
\bibfield{author}{\bibinfo{person}{Preethi Seshadri}, \bibinfo{person}{Sameer Singh}, {and} \bibinfo{person}{Yanai Elazar}.} \bibinfo{year}{2023}\natexlab{}.
\newblock \showarticletitle{The bias amplification paradox in text-to-image generation}.
\newblock \bibinfo{journal}{\emph{arXiv preprint arXiv:2308.00755}} (\bibinfo{year}{2023}).
\newblock


\bibitem[Venkit et~al\mbox{.}(2023c)]%
        {venkit2023sentiment}
\bibfield{author}{\bibinfo{person}{Pranav Venkit}, \bibinfo{person}{Mukund Srinath}, \bibinfo{person}{Sanjana Gautam}, \bibinfo{person}{Saranya Venkatraman}, \bibinfo{person}{Vipul Gupta}, \bibinfo{person}{Rebecca~J Passonneau}, {and} \bibinfo{person}{Shomir Wilson}.} \bibinfo{year}{2023}\natexlab{c}.
\newblock \showarticletitle{The Sentiment Problem: A Critical Survey towards Deconstructing Sentiment Analysis}. In \bibinfo{booktitle}{\emph{Proceedings of the 2023 Conference on Empirical Methods in Natural Language Processing}}. \bibinfo{pages}{13743--13763}.
\newblock


\bibitem[Venkit et~al\mbox{.}(2023a)]%
        {venkit2023nationality}
\bibfield{author}{\bibinfo{person}{Pranav~Narayanan Venkit}, \bibinfo{person}{Sanjana Gautam}, \bibinfo{person}{Ruchi Panchanadikar}, \bibinfo{person}{Ting-Hao Huang}, {and} \bibinfo{person}{Shomir Wilson}.} \bibinfo{year}{2023}\natexlab{a}.
\newblock \showarticletitle{Nationality Bias in Text Generation}. In \bibinfo{booktitle}{\emph{Proceedings of the 17th Conference of the European Chapter of the Association for Computational Linguistics}}. \bibinfo{pages}{116--122}.
\newblock


\bibitem[Venkit et~al\mbox{.}(2022)]%
        {venkit2022study}
\bibfield{author}{\bibinfo{person}{Pranav~Narayanan Venkit}, \bibinfo{person}{Mukund Srinath}, {and} \bibinfo{person}{Shomir Wilson}.} \bibinfo{year}{2022}\natexlab{}.
\newblock \showarticletitle{A study of implicit bias in pretrained language models against people with disabilities}. In \bibinfo{booktitle}{\emph{Proceedings of the 29th International Conference on Computational Linguistics}}. \bibinfo{pages}{1324--1332}.
\newblock


\bibitem[Venkit et~al\mbox{.}(2023b)]%
        {venkit2023automated}
\bibfield{author}{\bibinfo{person}{Pranav~Narayanan Venkit}, \bibinfo{person}{Mukund Srinath}, {and} \bibinfo{person}{Shomir Wilson}.} \bibinfo{year}{2023}\natexlab{b}.
\newblock \showarticletitle{Automated Ableism: An Exploration of Explicit Disability Biases in Sentiment and Toxicity Analysis Models}. In \bibinfo{booktitle}{\emph{Proceedings of the 3rd Workshop on Trustworthy Natural Language Processing (TrustNLP 2023)}}. \bibinfo{pages}{26--34}.
\newblock


\bibitem[Xu et~al\mbox{.}(2023)]%
        {xu2023combating}
\bibfield{author}{\bibinfo{person}{Danni Xu}, \bibinfo{person}{Shaojing Fan}, {and} \bibinfo{person}{Mohan Kankanhalli}.} \bibinfo{year}{2023}\natexlab{}.
\newblock \showarticletitle{Combating Misinformation in the Era of Generative AI Models}. In \bibinfo{booktitle}{\emph{Proceedings of the 31st ACM International Conference on Multimedia}}. \bibinfo{pages}{9291--9298}.
\newblock


\bibitem[Zack et~al\mbox{.}(2023)]%
        {zack2023coding}
\bibfield{author}{\bibinfo{person}{Travis Zack}, \bibinfo{person}{Eric Lehman}, \bibinfo{person}{Mirac Suzgun}, \bibinfo{person}{Jorge~A Rodriguez}, \bibinfo{person}{Leo~Anthony Celi}, \bibinfo{person}{Judy Gichoya}, \bibinfo{person}{Dan Jurafsky}, \bibinfo{person}{Peter Szolovits}, \bibinfo{person}{David~W Bates}, \bibinfo{person}{Raja-Elie~E Abdulnour}, {et~al\mbox{.}}} \bibinfo{year}{2023}\natexlab{}.
\newblock \showarticletitle{Coding Inequity: Assessing GPT-4's Potential for Perpetuating Racial and Gender Biases in Healthcare}.
\newblock \bibinfo{journal}{\emph{medRxiv}} (\bibinfo{year}{2023}), \bibinfo{pages}{2023--07}.
\newblock


\end{thebibliography}

\appendix

\end{document}